\documentclass[9pt,twocolumn,twoside]{extarticle}

\usepackage{times}
\usepackage[T1]{fontenc}
\usepackage[colorlinks=true,linkcolor=black,citecolor=blue,urlcolor=blue]{hyperref}
\usepackage{graphicx}
\usepackage{booktabs}
\usepackage{rotating} 
\usepackage{placeins} 
\usepackage{amsmath} 
\usepackage{setspace} 
\usepackage{caption}  
\usepackage{geometry}
\newcommand{\checkref}[1]{\textcolor{red}{#1}}

\title{Social balance in directed networks}

\author
{Bingjie Hao$^{1}$, Elijah Platnick$^{1}$, István A. Kovács,$^{1,2,3,\ast}$\\
\\
\footnotesize{$^{1}$Department of Physics and Astronomy, Northwestern University, Evanston, IL 60208, USA}\\
\footnotesize{$^{2}$Northwestern Institute on Complex Systems, Northwestern University, Evanston, IL 60208, USA}\\
\footnotesize{$^{3}$Department of Engineering Sciences and Applied Mathematics, Northwestern University, Evanston, IL 60208, USA}\\
\footnotesize{$^\ast$To whom correspondence should be addressed; E-mail:  \href{mailto:istvan.kovacs@northwestern.edu}{istvan.kovacs@northwestern.edu}}
}

\begin{document}
\maketitle

\begin{abstract} 
Social networks inherently exhibit complex relationships that can be positive or negative, as well as directional. Understanding balance in these networks is crucial for unraveling social dynamics, yet traditional theories struggle to incorporate directed interactions. This perspective presents a comprehensive roadmap for understanding balance in signed directed networks, extending traditional balance theory to account for directed interactions. Balance is indicated by the enrichment of higher-order patterns like triads compared to an adequate null model, where the network is randomized with some key aspects being preserved. 
Finding appropriate null models has been a challenging task even without considering directionality, which largely expands the space of potential null models. Recently, it has been shown that in the undirected case both the network topology and the signed degrees serve as key factors to preserve.
Therefore, we introduce a maximally constrained null model that preserves the directed topology as well as node-level features given by the signed unidirectional, reciprocated, and conflicting node degrees. 
Our null model is based on the maximum-entropy principle and reveals consistent patterns across large-scale social networks. We also consider directed generalizations of balance theory and find that the observed patterns are well aligned with two proposed directed notions of strong balance. 
Our approach not only unveils balance in signed directed networks but can also serve as a starting point towards generative models of signed directed social networks, advancing our understanding of complex social systems and their dynamics.
\end{abstract}



\section*{Introduction}
Understanding the patterns of large-scale social networks has been a long-standing challenge. It happened only recently that a clear sign of strong structural balance~\cite{haoProperNetworkRandomization2024} has been observed consistently in multiple large-scale social networks \cite{haoProperNetworkRandomization2024,galloTestingStructuralBalance2024}. At least without considering directionality, the theory of strong balance dictates that both the friend of a friend, and the enemy of an enemy is expected to be a friend more often than expected by chance. On the other hand, the enemy of a friend or the friend of an enemy is often an enemy. In other words, an undirected triad is balanced if it contains an even number of negative links (zero or two). While many triads are unbalanced in real networks, balanced (imbalanced) triangles happen more (less) often than expected ``by chance'' in a randomized network ensemble given by an appropriate null model. 

Clearly, null models play a crucial role in network analyses by providing a baseline for comparing observed network structures against random expectations. 
The primary use of null models in this context is to differentiate between statistically significant patterns and those that could arise by chance given certain node-level constraints.
The most basic null model 
is a random graph with the same number of nodes and links as the observed network, for example given by the 
Erdős–Rényi model~\cite{bollobasRandomGraphs2001}. For signed networks, this basic model can be extended by randomly assigning positive or negative signs to links with a probability that maintains the overall sign ratio of the empirical network~\cite{szellMultirelationalOrganizationLargescale2010}.
While such simple null models preserve global network properties such as density and sign ratio, they fail to capture more detailed structural features. 
Consequently, when comparing such simplistic null models to empirical social networks, the interpretation of the results often becomes problematic or misleading. Specifically, the observed differences between the null model and the real network conflate multiple factors, blurring the true patterns of social networks~\cite{haoProperNetworkRandomization2024}.

The recent confirmation of balance in the undirected case required methodological advances in capturing key constraints in the null models. 
Both preserving the network topology and the individual sign preferences has been identified as key ingredients for such null models \cite{haoProperNetworkRandomization2024,galloTestingStructuralBalance2024}. 
By preserving the network topology, we separate factors that dictate which links are formed from those that determine the sign of each link, the latter being the subject of questions related to social balance.
The network topology is resulting from manifold physical and social constraints on forming connections. For instance, geographic proximity often limits potential interactions~\cite{preciadoDoesProximityMatter2012,kaboProximityEffectsDynamics2014,stopczynskiHowPhysicalProximity2018}, while organizational hierarchies can restrict communication pathways~\cite{gupteFindingHierarchyDirected2011,redheadSocialHierarchiesSocial2022}. Even in online social networks, where physical distance is less of a barrier, constraints persist in the form of language barriers, cultural differences, and algorithmically curated content exposure~\cite{bakshyExposureIdeologicallyDiverse2015,dunbarOnlineSocialMedia2016}. These digital constraints create virtual ``neighborhoods'' that shape connection patterns in ways analogous to physical proximity in offline networks.
If these topological constraints are relaxed in a null model, the resulting randomized networks may exhibit unrealistic connection patterns, leading to an overestimation of the significance of the observed network features.

Additionally, signed node degree in social networks represents the capacity or willingness of individuals to maintain positive or negative connections, which may subject to cognitive and time limitations~\cite{miritelloLimitedCommunicationCapacity2013,ciottiDegreeCorrelationsSigned2015a}. 
A null model without constraining the signed degree of individual nodes ignores the reality that every node is not equally friendly or antagonistic \cite{haoProperNetworkRandomization2024}.

Yet, even in these state-of-the-art studies \cite{haoProperNetworkRandomization2024,galloTestingStructuralBalance2024} a key aspect of the interactions has been ignored, namely that interactions can often happen in one direction only. 
From the perspective of a node, connections can be viewed as incoming and outgoing connections. Outgoing connections represent sociability while incoming connections may indicate the popularity of an entity. The degrees associated with these two types of connections-- in-degree and out-degree-- are not necessarily correlated. For example, while the Bitcoin-Alpha network shows a strong correlation between in- and out-degrees, the Slashdot shows a much weaker correlation, highlighting the importance of directionality (Figure~\ref{fig:network}A).
Moreover, our friends may not always think about us as their friends, and similarly for our foes. While the reported fraction of reciprocated interactions depends on the dataset~\cite{galloPatternsLinkReciprocity2024}, a large fraction of relationships can be non-reciprocated (Table~\ref{tab:dataset_overview} and Figure \ref{fig:network}B), also depending on whether they were positive (trust/friends) or negative (distrust/enemies). For instance, the Bitcoin-Alpha and Bitcoin-OTC datasets exhibit approximately $80\%$ reciprocated links, whereas in the other three datasets examined, over half of the links are unidirectional.
Although it occurs relatively rarely (less than $2\%$ of total links), our friend might also consider us as a foe, leading to a ``conflicting'' relationship.
Such observations indicate that directionality could play a substantial role in shaping higher-order network patterns.

While there is now a well-established notion of balance in undirected networks --- aligned with everyday intuition ---, it is less clear what the correct directed generalization would look like. Here, we aim to present a comprehensive roadmap towards understanding balance in signed directed networks, outlining the key steps to be taken and the current challenges.

\begin{figure}[htpb]
\centering
\includegraphics[width=0.9\linewidth]{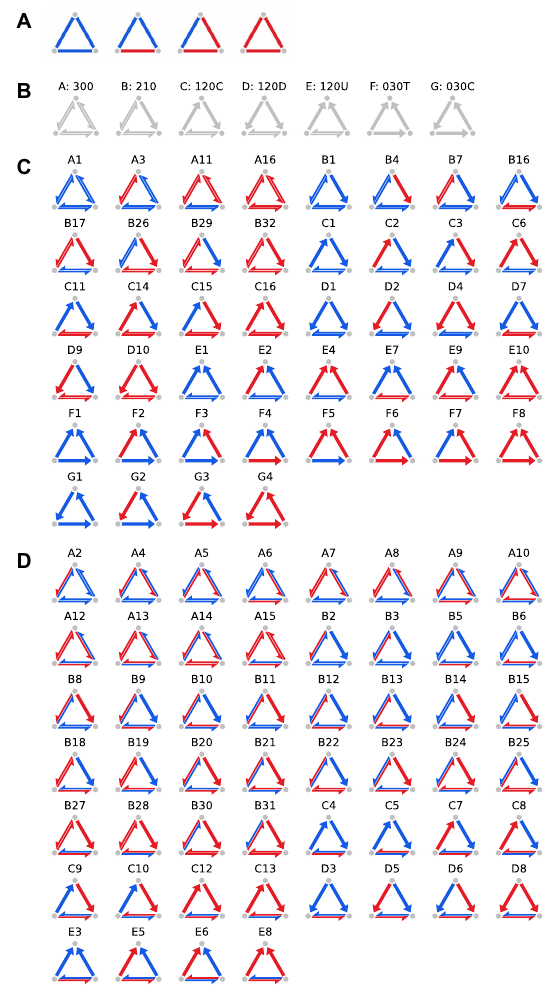}
\caption{Triad configurations. 
(A) Signed triads without directionality. (B) Directed triads without signs. The alphabet is used to distinguish between different topologies as defined in Ref.~\cite{hollandMethodDetectingStructure1970}.  (C) Triads without conflicting links: All links within these triads are either consistently positive (blue), negative (red), or unidirectional. (D) Triads with conflicting links: These triads contain at least one pair of mutual links with conflicting signs (one positive, one negative). Each triad is labeled with an alphanumeric code denoting distinct topological categories in (B). }
\label{fig:triads}
\end{figure}

As a starting point, signed directed networks pose manifold challenges compared to signed undirected networks or unsigned directed networks, namely: 
i) More link types: unidirectional \(A \rightarrow B\) ($+$ or $-$); reciprocated \(A \leftrightarrow B\), meaning that the connection exists in both directions with the same sign; and conflicting, meaning that the sign of the two directions differs.
ii) More primary node features: signed unidirectional degree ($k^{in_u \pm}, k^{out_u \pm}$); signed reciprocated degree ($k^{r \pm}$); signed conflicting degree ($k^{c \pm}$, where the indicated sign corresponds to the outgoing arrow). 
iii) A combinatorial increase in secondary node features, derived from primary features, like the total positive (negative) degree, the total outgoing degree, or the total outgoing positive degree, etc.
iv) A drastic increase in the potential network patterns (96 configurations) at the level of fully connected three-node triads, compared to the undirected signed case (4 configurations) and the unsigned directed case (7 configurations), see Figure~\ref{fig:triads}. As 
the presence of conflicting links might 
already violate classical balance theory assumptions \cite{heiderAttitudesCognitiveOrganization1946}, 
we consider triads without (Figure \ref{fig:triads}B) and with (Figure \ref{fig:triads}C) conflicting links separately.
v) A proliferation of potential null models depending on which network features are preserved.
vi) A broad range of (largely unexplored) potential extensions and alternatives of balance theory for directed networks.

Most importantly, the choice of null models plays a crucial role in assessing the significance of the observed balance in signed directed networks. Just like in the undirected case, each choice of the preserved network features can lead to a different interpretation of the network structure. 
Two fundamental properties to consider are the network topology and some variation of (signed and/or directed) node degrees. The decision to preserve network topology depends on the underlying assumptions about the system being studied. Allowing disruption of the network topology in the null model assumes that all connections can be potentially established, which is often unrealistic in many real-world systems. For example, in a product-competition network~\cite{cuiWeightedStatisticalNetwork2022}, customer choices are restricted to products within their consideration set, forming an underlying network structure of constraints. If we had the constraint data on which connections are allowed to form, we could randomize the topology of the network within the allowed subspace~\cite{kovacsUncoveringGeneticBlueprint2020}. 
For instance, in brain networks, only neurons in physical contact can form synapses to communicate with each other~\cite{cookNeuronalContactPredicts2023}, imposing inherent contact constraints on possible connections. 
However, we often do not have access to such detailed information for social networks, leaving two options: either allowing all potential connections or freezing the topology completely.


In the undirected case, the preservation or randomization of the topology and the signed node degrees yields $2\times2=4$ potential null models. A natural starting point for the directed case, --- and our main focus ---, is the maximally constrained null model that preserves the topology and all primary (and therefore all secondary) node degrees. 
%
%
Note, however, that the landscape of potential null models for directed networks is considerably more complex than in the undirected case. As we mentioned, preserving any consistent subset of the primary and secondary node degrees is a potential mathematical option. While in the undirected case it was possible to follow a step-by-step elimination process until the maximally constrained null model was the only reliable option left \cite{haoProperNetworkRandomization2024}, carrying out a similar process appears to be impractical in the directed case. That means that there could be multiple viable alternative null models to explore. Yet, the most constrained version serves as a good starting point. If we still see a difference compared to this null model, then the datasets clearly have some sign patterns that remained unexplained by the local node-level features or the topology itself. A less restrictive null model, the ``signed directed'' null model, only preserves the topology and the signed in- and out- degrees, corresponding to the assumption that all nodes are equally likely to form mutual links.
We present the results for both null models in this perspective to demonstrate the importance of choosing null models to unravel the true patterns of signed directed networks.




We consider five large-scale signed directed social networks from various fields: (a) Bitcoin-Alpha: a who-trusts-whom network of people who trade using Bitcoin on a platform called Bitcoin Alpha \cite{kumarREV2FraudulentUser2018,kumarEdgeWeightPrediction2016}; (b) Bitcoin-OTC: a who-trusts-whom network of people who trade using Bitcoin on a platform called Bitcoin OTC \cite{kumarREV2FraudulentUser2018,kumarEdgeWeightPrediction2016}; (c) Slashdot: a friend-or-foes network between users of a technology-related news website called Slashdot \cite{leskovecCommunityStructureLarge2008}; (d) Epinions: a who-trusts-whom online social network of a general consumer review site called Epinions \cite{leskovecSignedNetworksSocial2010}; (e) Pardus: an accumulated network of relationships among players of a Massive Multiplayer Online Game (MMOG) called Pardus \cite{szellMultirelationalOrganizationLargescale2010}. Details of these datasets can be found in the Supplementary Material and Table \ref{tab:dataset_overview}.

\begin{table*}[ht]
\centering
\begin{tabular}{lllp{0.7cm} p{0.7cm} lllll}
\toprule
      Dataset &   nodes &   links &  + ratio &  \text{--} ratio & + unidirectional & \text{--} unidirectional & + reciprocated & \text{--} reciprocated & conflicting \\
\midrule
Bitcoin\text{--}Alpha &   3,775 &  24,180 &     0.94 &     0.06 &      3,045 (13\%) &       1,015 (4\%) &   19,352 (80\%) &       272 (1\%) &    496 (2\%) \\
  Bitcoin\text{--}OTC &   5,875 &  35,587 &     0.90 &     0.10 &      4,794 (13\%) &       2,597 (7\%) &   26,872 (76\%) &       608 (2\%) &    716 (2\%) \\
     Epinions & 119,130 & 833,390 &     0.85 &     0.15 &    459,730 (55\%) &    115,414 (14\%) &  248,166 (30\%) &     4,690 (1\%) &  5,390 (1\%) \\
     Slashdot &  82,140 & 549,202 &     0.77 &     0.23 &    338,743 (62\%) &    113,017 (21\%) &   84,380 (15\%) &     9,164 (2\%) &  3,898 (1\%) \\
       Pardus &  12,740 & 167,924 &     0.55 &     0.45 &     28,698 (17\%) &     63,612 (38\%) &   63,060 (38\%) &    10,350 (6\%) &  2,204 (1\%) \\
\bottomrule
\end{tabular}
\caption{Overview of the empirical social networks. Columns show dataset name, node count, link count, positive and negative link ratios, and counts (with percentages) of +/\text{--} unidirectional, +/\text{--} reciprocated, and conflicting links. The percentages represent the relative frequency of each link type as a proportion of the total number of links in the network. }
\label{tab:dataset_overview}
\end{table*}



\section*{The spectrum of null models}

\begin{figure*}[htpb]
\centering
\includegraphics[width=.9\linewidth]{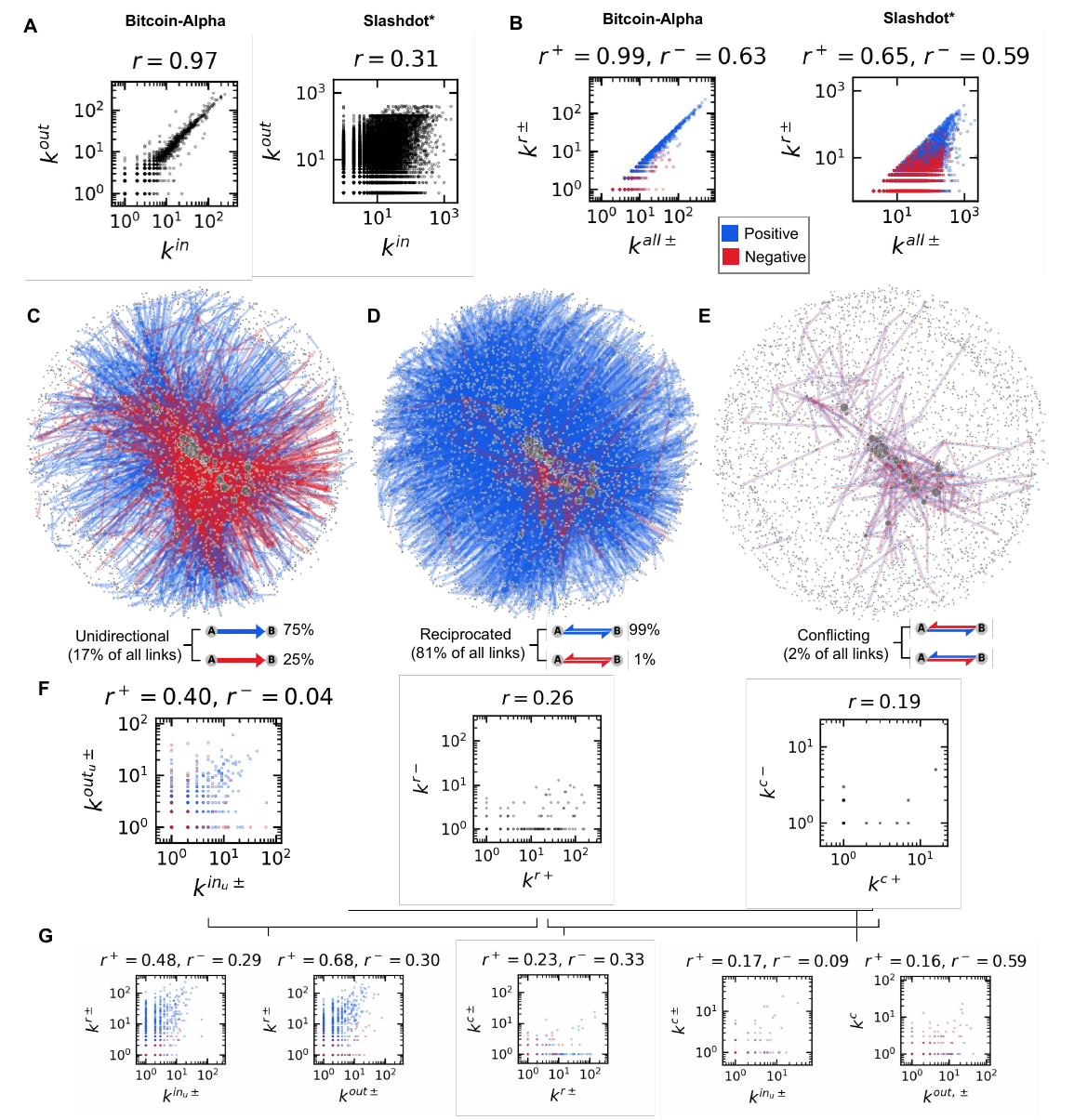}
\caption{Overview of distinct link types and degree correlations in empirical networks.
(A) The correlation between in- and out-degree ($k^{in}, k^{out}$) of Bitcoin-Alpha and Slashdot networks.
(B) The correlation between signed total node degree ($k^{all \pm}$) and signed reciprocated degree ($k^{r \pm}$) of Bitcoin-Alpha and Slashdot networks.
(C-E) The Bitcoin-Alpha network decomposed into unidirectional, reciprocated and conflicting links. The percentage of each link type relative to total links and their positive/negative composition are shown below each network layout.
Negative links (red) are overlaid on positive links (blue), with node sizes proportional to the total degree. Node positions are computed using a spring-embedded layout algorithm~\cite{kamadaAlgorithmDrawingGeneral1989} considering all links regardless of their signs.
(F) Correlation plots of the primary node degrees associated with each link type in panels C-E.
(G) Correlation plots between degrees correspond to different link types.
All correlation plots use logarithmic scales to accommodate the broad range of degree values, though correlation coefficients ($r, r^{+}, r^{-}$) are computed using linear-scale degree values. Throughout all panels, blue and red consistently represent positive and negative links/degrees, respectively. $ ^{\ast}$Slashdot limits the total number of connections to 200 (or 400 for subscribers)~\cite{kunegisSlashdotZooMining2009}.
}
\label{fig:network}
\end{figure*}

Network structure is shaped by node-level preferences as well as pairwise or higher-order wiring mechanisms. A key step towards unveiling the wiring mechanisms is to compare with null models that match the key features of individual nodes.
Formally, a null model is an ensemble of random graphs that is constrained by some selected features of the original network. 
In directed networks, it is natural to consider the signed in- and out-degrees separately since they might originate from different mechanisms. For instance, the positive in-degree can indicate popularity or prestige, while the positive out-degree can result from sociability or influence-seeking behavior~\cite{chaMeasuringUserInfluence2010,srinivasIdentificationInfluentialNodes2015}. Thus, the in- and out-degrees are not necessarily well correlated in empirical networks as shown in Figure~\ref{fig:network}A. 

Reciprocity is another key aspect of social systems.
As shown in Figure~\ref{fig:network}B, for negative links, both the Bitcoin-Alpha and Slashdot networks exhibit moderate correlations between reciprocated degrees and total degrees, indicating heterogeneous reciprocity patterns across nodes. This heterogeneity suggests that nodes with similar total negative degrees may display varying tendencies to engage in mutual negative relationships. 
For positive links, reciprocity can be either highly homogeneous (Bitcoin-Alpha) or not (Slashdot), depending on the dataset. Although conflicting links typically account for less than $2\%$ of total links, we observed that the tendency to form conflicting links can be highly heterogeneous among nodes (Figure \checkref{S1}).
We further check the correlations between all six independent primary node degrees and find that Pearson correlation coefficients are mostly below 0.50, with a few exceptions that can go up to 0.77 (Figure~\ref{fig:network}F,G and Table \checkref{S1-5}).
These weak to moderate correlations likely suggest that each link type is influenced differently by social processes.
For example, reciprocated positive links may represent mutual friendship, while conflicting links could indicate complex status relationships \cite{leskovecSignedNetworksSocial2010}. 
Critically, these correlations demonstrate that constraining one type of degree does not automatically constrain others. For example, maintaining the positive unidirectional in-degree does not guarantee the preservation of the positive reciprocated degree. Hence providing motivation for the maximally constrained null model that preserves each primary degree separately. However, note that constraining the network topology makes some primary degrees depend on other degrees. For example, fixing the topology preserves both in- and out-degrees for unidirectional links. Consequently, if the positive unidirectional in-degree is preserved, the corresponding negative unidirectional out-degree is also automatically preserved. A similar relationship exists for signed reciprocated and conflicting degrees, which reduces the independent degrees considered in the null model, as described in detail in the Supplementary Material.

\begin{figure*}[htpb]
\centering
\includegraphics[width=.8\linewidth]{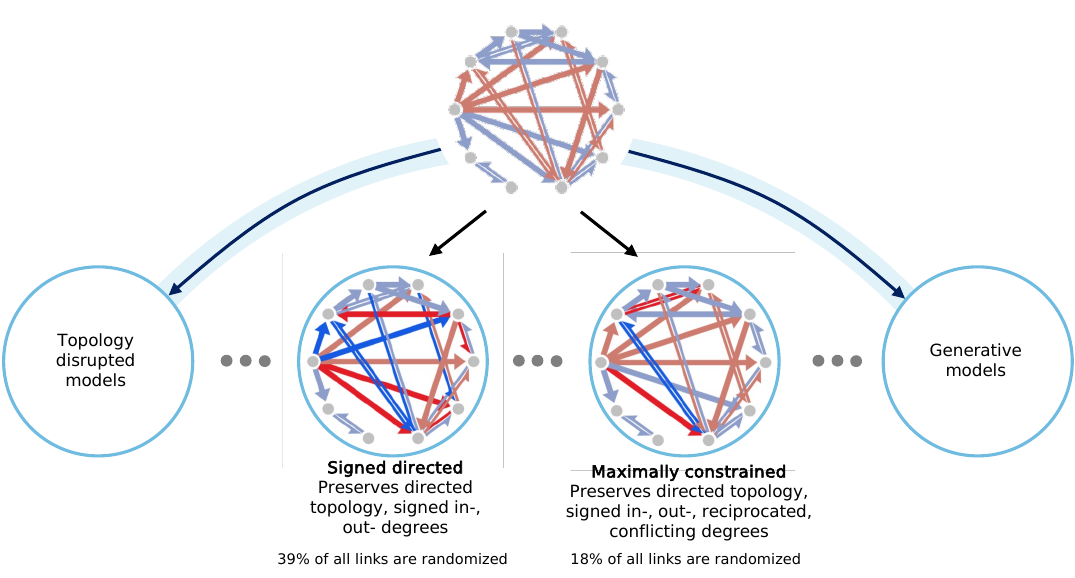}
\caption{Randomizing signed directed networks. The diagram illustrates a progression from topology-disrupted models (left) to generative models (right), with intermediate models. The top network represents the original structure, with positive links in blue and negative links in red. Two example null models are shown: (1) the signed directed model that preserves directed topology and signed in- and out-degrees ($39\%$ randomized), and (2) the maximally constrained model that preserves directed topology with signed in-, out-, reciprocated-, and conflicting-degrees ($18\%$ randomized). Darker colors highlight randomized links. 
}
\label{fig:null_model}
\end{figure*}

On the way to the maximally constrained null model, as we incorporate more constraints into the null model, the fraction of the network that is being randomized decreases (Figure \ref{fig:null_model}).
While the maximally constrained null model matches all node-level features, in principle, we could include additional constraints based on pairwise and higher-order network properties. Such a non-local null model would eventually capture all characteristics of the empirical social networks. At this point, the non-local null model could serve as a 
generative model, capable of producing synthetic networks statistically indistinguishable from the observed network across multiple measures. In this sense, the progression from null models to generative models represents a continuum of increasing structural fidelity, reflecting our evolving understanding of the fundamental organizing principles in social networks.

However, as we mentioned above, for the purpose of detecting the patterns emerging from wiring mechanisms, we only want to include local constraints at the level of individual nodes.
To sum, in this perspective we present two null models to illustrate how incorporating appropriate constraints reveals the hidden structure of empirical social networks. Based on a previous study \cite{haoProperNetworkRandomization2024}, we consider both network topology and signed degrees as fundamental constraints. We extend this approach to signed directed networks by preserving directed topology and signed in- and out-degrees (signed directed null model). This null model is efficiently generated by applying maximum-entropy randomization to directed networks, maintaining average in- and out-degrees across the ensemble (see Supplementary Material for details).
Alternatively, the maximally constrained null model considers all primary node degrees from three distinct types of directed links: unidirectional, reciprocated, and conflicting. In addition to directed topology, the maximally constrained null model preserves the signed in-, out-, reciprocated-, and conflicting-degree of each node, when averaged over the ensemble of the null model. We implement this model by decomposing the directed network into three independent subgraphs: (1) unidirectional positive and negative links, (2) reciprocated positive and negative links, and (3) conflicting links. Maximum-entropy randomization is applied to each subgraph separately. The union of the links in the resulting randomized subgraphs provides 
the complete null model, as formulated in \checkref{Section 3} of the Supplementary Material. 

\section*{Directed notions of balance} 
\begin{figure}
\centering
\includegraphics[width=.9\linewidth]{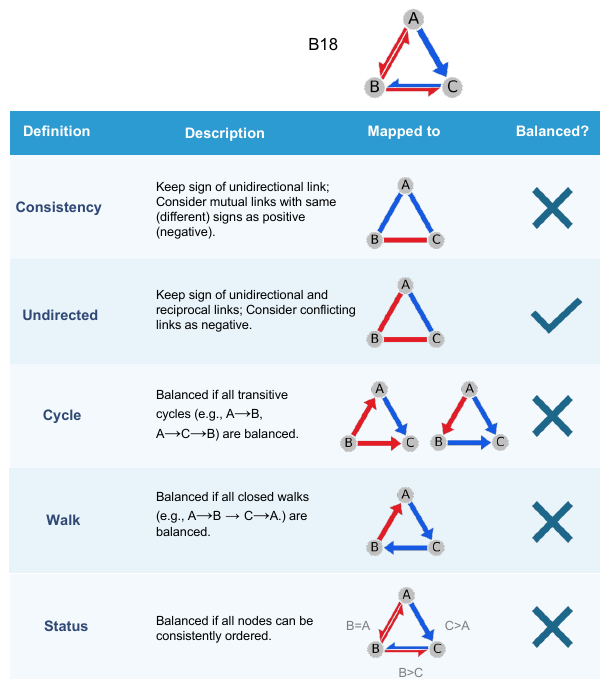}
\caption{Definitions of balance for signed directed triads. An example triad is considered under different definitions of balance. Positive links are shown in blue and negative links are shown in red. }
\label{fig:balance_def}
\end{figure}

Extending the definition of balance to directed network necessitates understanding the role played by reciprocated 
and conflicting links. 
For example, negative reciprocated links can either be considered as equivalent to a negative link in the undirected case, or be interpreted as a sign of balance, due to sign concordance~\cite{galloPatternsLinkReciprocity2024}.
However, as in undirected networks, a definition of balance should not be based on isolated links as such an approach ignores the complex interdependencies of social systems. Instead, a comprehensive definition of balance should consider the dynamics among a group of entities, resulting in patterns beyond those given by individual node-level features. 
Here, we discuss several definitions of strong balance based on (fully connected) triads in directed networks, extending balance theory from the undirected case to the directed case, as illustrated in 
Figure \ref{fig:balance_def}. 


As direct extensions of undirected balance theory, we first propose two definitions that transform directed links into undirected links.
The \textit{Undirected} definition simplifies the network by converting reciprocated $+/+$ to positive and $-/-$ to negative links, while treating conflicting $+/-$ as negative. In contrast, the \textit{Consistency} definition focuses on sign agreement, treating both $+/+$ and $-/-$ as positive, with only $+/-$ considered negative. This distinction depends on the interpretation of reciprocated negative links ($-/-$): the \textit{Undirected} approach views them as elevated discord, while the \textit{Consistency} approach sees them as a form of consistency in negative sentiment. Both definitions consider unidirectional links as undirected and preserve their signs.

As an alternative, we build upon the observation that the pattern $A\rightarrow B\rightarrow C\leftarrow A$, known as a transitive cycle~\cite{arefMultilevelStructuralEvaluation2020}, frequently appears in most considered datasets ($>30\%$ of total triads in Epinions, Slashdot, Pardus). Thus, we consider a definition of balance based on transitive cycles (\textit{Cycle}). This definition considers the consistency of sentiment between the direct interaction from $A$ to $C$ and the indirect interaction from $A$ to $C$ through $B$.
We consider a cycle as balanced if it has an even number of negative links, and unbalanced otherwise. A triad is considered balanced only if all transitive cycles are balanced. 

Another definition of directed balance considers the closed walks involving all three triad nodes (\textit{Walk}). Closed walks represent paths where information or influence can flow back to its origin. This circular flow within a triad can reinforce or counteract itself, depending on the signs of the links. We consider closed walks that encompass all nodes without repeating nodes and consider them as balanced if a given walk contains an even number of negative links, and unbalanced otherwise. A triad is considered balanced only if all closed walks are balanced. 

The final definition we consider is grounded in status theory \cite{guhaPropagationTrustDistrust2004,leskovecSignedNetworksSocial2010}, which offers a distinct perspective on balance in signed directed networks, particularly relevant in hierarchical social structures. According to status theory, the sign of a link between two nodes is determined by the perceived difference in their social status (\textit{Status}). 
Specifically, a positive link from node $A$ to node $B$ indicates that $A$ perceives $B$ as having a higher status, while a negative link suggests that $A$ views $B$ as having a lower status. In this definition, balance is achieved when all three nodes of a triad can be placed in a consistent status order. 
While previous studies of status theory have focused less on reciprocated links, real social systems often contain reciprocated links, potentially corresponding to 
equal status. Thus, here we introduce an extended notion of status theory by considering reciprocated positive or negative links between two nodes as indicators of having equal status. For example, if $A$ positively links to $B$ and $B$ also positively links to $A$, such triad is considered as balanced as long as both $A$ and $B$ have higher, lower, or equal status relative to $C$. 


Note that these prospective definitions of balance already differ even at the level of fully reciprocated configuration with consistent signs ($A1,A11,A3,A16$ in Figure~\ref{fig:results}). In these cases, according to the \textit{Undirected} definition, directionality plays no role, as mutual links with identical signs can be considered as undirected links without loss of information. 
On the contrary, both the \textit{Consistency} and \textit{Status} definitions suggest that triads $A3$ and $A16$ should be balanced, while the \textit{Undirected} definition indicates they should be unbalanced (Figure \ref{fig:results}). This discrepancy indicates that the \textit{Consistency} and \textit{Status} definitions should not be considered as extensions of the undirected notion of balance. Instead, these definitions may offer complementary insights into signed directed networks if they demonstrate consistency with empirical data.

\section*{Balance is observed in directed social networks}

To quantitatively understand the balance in signed directed networks, we consider all signed and directed triads presented in the empirical social systems. 
As a standard measure, we use the $z$-score  to quantify how the observed triad frequencies deviate from the null models. The $z$-score is calculated as
\begin{equation}
z = \frac{f_{obs}-\langle f_{null} \rangle}{\sqrt{\sigma_{obs}^2+\sigma_{null}^2}},
\end{equation}
where $f_{obs}$ is the observed frequency of a given triad and $\langle f_{null} \rangle$ is the mean frequency of the same triad type averaged over 1000 independently generated null model networks. The denominator represents the total uncertainty, combining two sources of uncertainty: $\sigma_{null}$, the standard deviation of triad frequencies across all null model samples, and $\sigma_{obs}$, the estimated shot noise $\sigma_{obs}\approx\sqrt{f_{obs}}$, assuming a Poisson distribution for the occurrence of each triad type.
This approach allows us to account for both the variability in the null model and the inherent statistical fluctuations in the observed network, providing a robust measure of the significance of triad frequency deviations. A triad is considered as significantly overrepresented when $z>2$ and significantly underrepresented when $z<-2$. 
Any $|z|<2$ score means that 
the triad does not deviate substantially from the null model. 

\begin{figure*}[htpb]
\centering
\includegraphics[width=\linewidth]{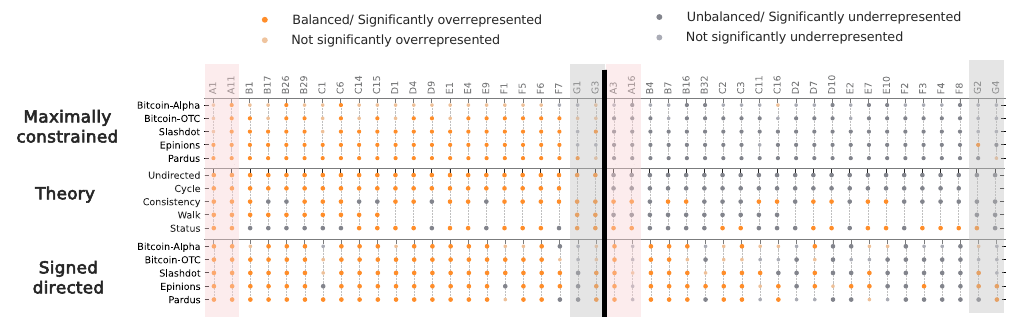}
\caption{Comparison of the observed triad statistics to null models and balance theories for triads without conflicting links. 
Triads are ordered based on their expected balance according to the \textit{Undirected} definition, with balanced triads presented first, followed by a black line and then unbalanced triads.
The ``Maximally constrained'' and ``Signed directed'' rows show $z$-scores quantifying the deviation of triad frequencies in empirical social networks from corresponding null model expectations. Orange (gray) dots indicate significant overrepresentation (underrepresentation) with $z>2$ ($z<-2$), while lighter colors indicate results with ($|z|<=2$). The ``Theory'' rows indicate whether each triad configuration is classified as balanced (orange) or unbalanced (gray) according to a certain definition of balance. 
The spot is left blank if the theory is inconclusive about the balance of that triad. 
Triads that contain only reciprocal links are shaded in red. Inconclusive triads without significant results are shaded in gray.
}
\label{fig:results}
\end{figure*}

First, we consider the maximally constrained null model, which preserves topology and all primary node degrees.
For each triad, we note consistency with a given definition of balance 
if there are no statistically significant contradicting conclusions regarding over- or under-representation across all empirical datasets.
In Figure~\ref{fig:results}, we ordered all triads without conflicting links based on whether they are balanced or not under the \textit{Undirected} definition. 
We observe consistent alignment with the \textit{Undirected} and \textit{Cycle} definitions of balance across all datasets examined. The exceptions are triads $G1-G4$, where the results are not significant in most datasets, indicating that these specific configurations occur at frequencies similar to what would be expected by chance, given the preserved network constraints.
Regarding triads with conflicting links, at least partially due to their small numbers, the majority lacks sufficient statistical significance to determine over- or underrepresentation compared to the null model, leading to no definite conclusions of balance (Figure \checkref{S2}). Another possible interpretation is that the maximally constrained null model already captures the statistics of triads with conflicting links. This means that once we account for the signed degrees of unidirectional, reciprocated and conflicting links of each node, the frequencies of triads containing conflicting links are found to be fully explained by these lower-order network properties.

On the contrary, when comparing observed frequencies to the signed directed null model, which preserves directed topology and signed in- and out-degrees, we do not observe a clear pattern that aligns with any proposed balance definition (Figure \ref{fig:results}). The most notable trend is the underrepresentation of all triads with conflicting links across datasets, with the sole exception of triad $A8$ in the Bitcoin-Alpha network (Figure \checkref{S2}). However, this observation does not reveal direct insights about balance in these networks. Instead, it is a consequence of the overrepresentation of reciprocated links and underrepresentation of conflicting links at the link level, as suggested by other studies \cite{galloPatternsLinkReciprocity2024}. This link-level pattern propagates to the triadic level, resulting in the observed underrepresentation of triads containing conflicting links and the overrepresentation of triads without conflicting links in most cases.

\section*{Conclusion and Further Directions}

This perspective outlines a roadmap for addressing the challenges in understanding signed directed networks and establishes foundational steps toward uncovering their inherent patterns.
By implementing a maximally constrained null model, we identified robust structural patterns that consistently emerge in empirical social networks.
Comparing empirical results with theoretical definitions helped us narrow down the studied extensions of balance theory to \textit{Undirected} and \textit{Cycle} definitions. 


It is worth noting that the \textit{Status} definition, while potentially applicable in certain social systems as indicated by previous studies \cite{galloPatternsLinkReciprocity2024,leskovecSignedNetworksSocial2010}, shows contradictory conclusions when compared to empirical results for most triads. 
Such contradiction suggests that status dynamics may not be as universally applicable in explaining signed directed network structures as previously thought.


Note that while this perspective provides a maximally constrained null model as a starting point for understanding structural balance in signed directed networks, alternative null models may be considered.
Specifically, a better null model would (1) yield statistics that more closely match those of the empirical dataset, resulting in smaller absolute $z$-scores, and (2) effectively remove structural balance when applied to a reference network that initially exhibits such balance, similar to the approach demonstrated in Ref. \cite{haoProperNetworkRandomization2024}. However, unlike undirected networks, constructing a suitable reference network for the directed case presents significant challenges, as discussed in the Introduction.

One limitation of the current study is that it focuses on triad patterns. However, to fully understand the organizing principles of social systems, or more broadly signed directed systems, it may be necessary to incorporate higher-order patterns~\cite{cosciaMultilayerGraphAssociation2021}, such as four-node patterns, as well.

Although our previous study~\cite{haoProperNetworkRandomization2024} found that square patterns also show balance in the undirected case, extending such analysis to directed networks may require a fundamentally different framework. The primary obstacle lies in the combinatorial explosion of possible configurations as the number of nodes increases. 
It is worth noting that analysis based on comparing patterns to null models inherently involves multiple hypothesis testing issues. In our analysis, we simultaneously test 96 variables that are not independent of each other. This approach can lead to an increased risk of Type I errors (false positives) if not properly addressed. For example, when testing multiple hypotheses at a certain significance level, the probability of observing at least one false positive result increases with the number of tests performed. Although in this study we examine our main results against several thresholds of $z$-scores (e.g., $|z| = 2$, $3$, $5$, and $10$, see \checkref{Figure S3}) and demonstrate robustness, future studies may benefit from more rigorous statistical approaches. For example, in addition to considering $z$-scores, a combination of fold change and empirical $p$-values may also be considered.

Moving forward along the roadmap, future studies could incorporate additional primary node features related to the node attributes, which is also known as node color~\cite{ribeiroDiscoveringColoredNetwork2014}. 
For instance, categorical node attributes (e.g., gender) could generate attribute-specific network properties, such as gender-specific degree, which quantifies the number of connections a node has to nodes of a particular gender category~\cite{psyllaRoleGenderSocial2017}. 
Similar to how distinct link types drastically increase the number of possible network patterns, the introduction of node colors multiplies the complexity of structural patterns that need to be properly considered.
Eventually, advancing towards generative models would require considering more node features and associated degrees, pairwise connections and even higher-order properties.

This work provides a roadmap towards understanding the formation of alliances and conflicts in various social contexts~\cite{rawlingsStructuralBalanceTheory2017,szellMultirelationalOrganizationLargescale2010}. Such a roadmap can be applied to a wide range of signed directed networks. 
For instance, applying our methods to \emph{C. elegans} neuronal networks~\cite{bentleyMultilayerConnectomeCaenorhabditis2016,harrisComputationalInferenceSynaptic2022,varshneyStructuralPropertiesCaenorhabditis2011}, which contain approximately 300 neurons with signed, directed synapses, could reveal whether principles of social balance have analogs in the neuronal organization. 
Incorporating higher-order or non-local constraints may eventually lead to an effective generative model that captures the organizational principles of neural networks with low-order properties~\cite{salovaCombinedTopologicalSpatial2024}. 




\textbf{Acknowledgment} We thank Michael Szell for providing the Pardus dataset and Kathleen Carrasco for useful comments and discussion.



\bibliography{ref}

\end{document}





\maketitle 

\tableofcontents
\newpage


\section{Correlation between distinct link types}

\begin{figure}[htpb]
\centering
\includegraphics[width=0.6\linewidth]{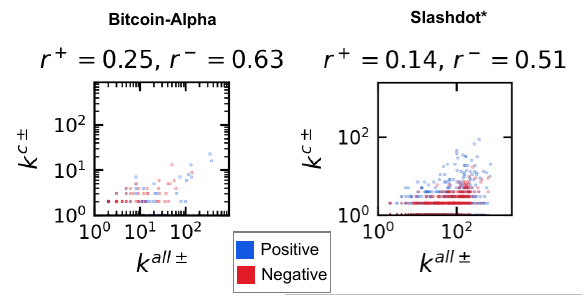}
\caption{The correlation between signed total node degree ($k^{all \pm}$) and signed conflicting degree ($k^{c \pm}$) of Bitcoin-Alpha and Slashdot networks. All correlation plots use logarithmic scales to accommodate the broad range of degree values,
though correlation coefficients ($r, r^+, r^-$) are computed using linear-scale degree values.}
\label{figS:conflict_inhomo}
\end{figure}

    Table~\ref{tab:corr_bitcoin-Alpha}-\ref{tab:corr_pardus} show Pearson correlation coefficients between each of the eight different edge types (+/- unidirectional in/out degree, +/- reciprocated degree, and conflicting degree split into + and - out degree). Each table gives the corresponding $r$ values for one of the five datasets.\\ \\

\begin{table}[h!]
    \centering
    \caption{Bitcoin Alpha Correlation Coefficients} 
    \scriptsize
    
    \begin{tabular}{l| c| c| c| c| c| c| c} 
         & $k^{in_u +}$ & $k^{out_u +}$ & $k^{in_u -}$ & $k^{out_u -}$ & $k^{r +}$ & $k^{r -}$ & $k^{c +}$\\  
        \midrule 
        $k^{out_u +}$    & 0.40 \\  %

        $k^{in_u -}$   & 0.13  & 0.13 \\
        $k^{out_u -}$    & 0.32    & 0.34    & 0.04 \\
        $k^{r +}$      & 0.48   & \textbf{0.68}  & 0.10 &0.35 \\
        $k^{r -}$      & 0.26 & 0.36    & 0.29 & 0.30 &0.26 \\
        
        $k^{c +}$      & 0.17    & 0.16    & 0.31 & 0.08 &0.23 &0.25\\
        $k^{c -}$     & 0.42 & 0.45    & 0.09 & \textbf{0.59} &\textbf{0.58} &0.33 &0.19\\

    \end{tabular}
\label{tab:corr_bitcoin-Alpha}
\end{table}

\begin{table}[h!]
    \centering
    \caption{Bitcoin OTC Correlation Coefficients} 
    \scriptsize
    
    \begin{tabular}{l| c| c| c| c| c| c| c} 
         & $k^{in_u +}$ & $k^{out_u +}$ & $k^{in_u -}$ & $k^{out_u -}$ & $k^{r +}$ & $k^{r -}$ & $k^{c +}$\\  
        \midrule 
        $k^{out_u +}$    & 0.30 \\  %

        $k^{in_u -}$   & 0.23   & 0.11 \\
        $k^{out_u -}$    & 0.32    & 0.23   & 0.12 \\
        $k^{r +}$      & 0.47    & \textbf{0.59}  & 0.11 & 0.34 \\
        $k^{r -}$      & 0.26 & 0.22    & 0.28 & 0.45 &0.31 \\
        
        $k^{c +}$      & 0.14    & 0.12    & 0.28 & 0.07 &0.23 &0.18\\
        $k^{c -}$     &0.38  & 0.31    & 0.14 & 0.49 &\textbf{0.62} &\textbf{0.54} &0.21\\

    \end{tabular}
\label{tab:corr_bitcoin-OTC}
\end{table}

\begin{table}[h!]
    \centering
    \caption{Slashdot Correlation Coefficients} 
    \scriptsize
    \begin{tabular}{l| c| c| c| c| c| c| c} 
         & $k^{in_u +}$ & $k^{out_u +}$ & $k^{in_u -}$ & $k^{out_u -}$ & $k^{r +}$ & $k^{r -}$ & $k^{c +}$\\  
        \midrule 
        $k^{out_u +}$    & 0.11 \\  %

        $k^{in_u -}$   &0.41   & 0.13 \\
        $k^{out_u -}$    & 0.06    & 0.32    & 0.13 \\
        $k^{r +}$      & 0.20  & 0.46  & 0.21 & 0.12 \\
        $k^{r -}$      & 0.08 & 0.19    & 0.19 & \textbf{0.51} &0.21 \\
        
        $k^{c +}$      & 0.05    & 0.09    & 0.13 & 0.01 &0.17 &0.03\\
        $k^{c -}$     & 0.08  & 0.19    & 0.13 & \textbf{0.55} &0.12 &0.39 &0.02\\
      
    \end{tabular}
\label{tab:corr_slash}
\end{table}

\begin{table}[h!]
    \centering
    \caption{Epinions Correlation Coefficients} 
    \scriptsize
    
    \begin{tabular}{l| c| c| c| c| c| c| c} 

         & $k^{in_u +}$ & $k^{out_u +}$ & $k^{in_u -}$ & $k^{out_u -}$ & $k^{r +}$ & $k^{r -}$ & $k^{c +}$\\  
        \midrule 
        $k^{out_u +}$    & 0.18 \\  %

        $k^{in_u -}$   & 0.33   & 0.13 \\
        $k^{out_u -}$    & 0.38   & 0.09    & 0.21 \\
        $k^{r +}$      & \textbf{0.52}   & 0.48  & 0.25 & 0.28 \\
        $k^{r -}$      & 0.34 & 0.07    & 0.33 & \textbf{0.70} &0.25 \\
        
        $k^{c +}$      & 0.07   & 0.35    &0.24 & 0.05 &0.15 &0.07\\
        $k^{c -}$     & 0.43  & 0.06    & 0.21 & \textbf{0.77} &0.28 &\textbf{0.70} &0.04\\

    \end{tabular}
\label{tab:corr_epinions}
\end{table}

\begin{table}[h!]
    \centering
    \caption{Pardus Correlation Coefficients} 
    \scriptsize
    
    \begin{tabular}{l| c| c| c| c| c| c| c} 
         & $k^{in_u +}$ & $k^{out_u +}$ & $k^{in_u -}$ & $k^{out_u -}$ & $k^{r +}$ & $k^{r -}$ & $k^{c +}$\\  
        \midrule 
        $k^{out_u +}$    & 0.33 \\  %

        $k^{in_u -}$   & 0.21   & 0.07 \\
        $k^{out_u -}$    & 0.33    & 0.34   & 0.11 \\
        $k^{r +}$      & \textbf{0.55}    & \textbf{0.65}  & 0.15 & 0.40 \\
        $k^{r -}$      &0.26 & 0.20    & 0.21 & 0.50 &0.31 \\
        
        $k^{c +}$      & 0.05    & 0.05   & 0.08 &0.03 
        &0.05 &0.04 \\
        $k^{c -}$     & 0.48  & 0.26    & 0.11 & 0.46 &0.39 &0.35  &0.03\\

    \end{tabular}
\label{tab:corr_pardus}
\end{table}

\section{Maximum-entropy based randomization}

\subsection{Randomization of unsigned directed network while keeping in-, out-, and reciprocated degrees}

In the main text, we use the terms "reciprocated" and "conflicting" to distinguish between mutual links with the same or different signs, respectively. For signed networks, we maintain this notation. In the unsigned case, however, we follow the convention of the field and use "reciprocated" to denote mutual links.

We denote the links in the directed network as 
\begin{equation}
    \sigma_{ij}=
    \begin{cases}
    1, & \text{if there is a link } j \to i \\
    0, & \text{otherwise}
    \end{cases}
\end{equation}

Ignoring the signs, each node in the networks has three degrees defined as
\begin{align}
k_i^{\text{in}} &= \sum_j \sigma_{ij} - k_r \\
k_i^{\text{out}} &= \sum_j \sigma_{ji} - k_r \\
k_i^r &= \sum_j \sigma_{ij} \sigma_{ji}
\end{align}
where $k_i^{\text{in}}$, $k_i^{\text{out}}$, $k_i^r$ represent the in-, out- and reciprocated degrees, respectively. The Hamiltonian can be written as
\begin{equation}
\begin{aligned}
H &= \sum_i \left( \theta_i^{\text{in}} k_i^{\text{in}} + \theta_i^{\text{out}} k_i^{\text{out}} + \theta_i^r k_i^r \right) \\
  &= \sum_i \left[ \theta_i^{\text{in}} \sum_j \sigma_{ij} - \theta_i^{\text{in}} \sum_j \sigma_{ij} \sigma_{ji} + \theta_i^{\text{out}} \sum_j \sigma_{ji} - \theta_i^{\text{out}} \sum_j \sigma_{ij} \sigma_{ji} + \theta_i^r \sum_j \sigma_{ij} \sigma_{ji} \right] \\
  &= \sum_{ij} \left( \theta_i^{\text{in}} + \theta_j^{\text{out}} \right) \sigma_{ij} - \sum_{ij} \left( \theta_i^{\text{in}} + \theta_j^{\text{out}} \right) \sigma_{ij} \sigma_{ji} + \sum_{ij} \theta_i^r \sigma_{ij} \sigma_{ji}.
\end{aligned}
\end{equation}
The partition function $Z=\sum_G e^{-H}$ becomes
\begin{equation}
\begin{aligned}
Z &= \sum_{\{ \sigma_{ij} \}} \prod_{ij} \exp \left[ - (\theta_i^{\text{in}} + \theta_j^{\text{out}}) \sigma_{ij} + (\theta_i^{\text{in}} + \theta_j^{\text{out}}) \sigma_{ij} \sigma_{ji} - \theta_i^r \sigma_{ij} \sigma_{ji} \right] \\
&= \prod_{ij} \sum_{\sigma_{ij} = 0}^{1} \exp \left[ - (\theta_i^{\text{in}} + \theta_j^{\text{out}}) \sigma_{ij} + (\theta_i^{\text{in}} + \theta_j^{\text{out}}) \sigma_{ij} \sigma_{ji} - \theta_i^r \sigma_{ij} \sigma_{ji} \right] \\
&= \prod_{ij} \left( 1 + \exp \left[ - (\theta_i^{\text{in}} + \theta_j^{\text{out}}) + (\theta_i^{\text{in}} + \theta_j^{\text{out}}) \sigma_{ji} - \theta_i^r \sigma_{ji} \right] \right).
\end{aligned}
\end{equation}
Denote the free energy $F = -\ln{Z}$, the node degrees can be calculated as
\begin{equation}
\begin{aligned}
\langle k_i^{\text{in}} \rangle &= \frac{\partial F}{\partial \theta_i^{\text{in}}}\\ 
&= \sum_j \frac{(1 - \sigma_{ji})}{1 + e^{ (\theta_i^{\text{in}} + \theta_j^{\text{out}}) - (\theta_i^{\text{in}} + \theta_j^{\text{out}}) \sigma_{ji} + \theta_i^r \sigma_{ji}}} \\
&= \sum_{ij (\sigma_{ji} = 0)} \frac{1}{1 + e^{\theta_i^{\text{in}} + \theta_j^{\text{out}}}} = \sum_{ij (\sigma_{ji} = 0)} \frac{1}{1 + \alpha_i^{in} \alpha_j^{out}},
\end{aligned}
\label{eq:k_i_in}
\end{equation}
where for simplicity we denote $\alpha_i^{in} = e^{\theta_i^{in}}, \alpha_i^{out} = e^{\theta_i^{out}}$. Similarly, the out- and reciprocated-degree can be calculated as
\begin{equation}
\begin{aligned}
\langle k_i^{\text{out}} \rangle &= \frac{\partial F}{\partial \theta_i^{\text{out}}}\\ 
&= \sum_{ij (\sigma_{ji} = 0)} \frac{1}{1 + \alpha_j^{in} \alpha_i^{out}},
\end{aligned}
\label{eq:k_i_out}
\end{equation}
\begin{equation}
\begin{aligned}
\langle k_i^{\text{r}} \rangle &= \frac{\partial F}{\partial \theta_i^{\text{r}}}\\ 
&= \sum_{ij (\sigma_{ji} = 1)} \frac{1}{1 + \alpha_i^{r}},
\end{aligned}
\label{eq:k_r}
\end{equation}
where we denote $\alpha_i^{r} = e^{\theta_i^{r}}$. These parameters $\alpha_i^{in}$, $\alpha_i^{out}$, $\alpha_i^{r}$ can be solved iteratively. The probability of having a unidirectional link between $i$ and $j$ is thus
\begin{equation}
    p_{ij}^{u} = \frac{1}{1 + \alpha^{in}_i \alpha^{out}_j},
\label{eq:uni_p}
\end{equation}
while the probability of having a reciprocated link between $i$ and $j$ is thus 
\begin{equation}
    p_{ij}^{r} = \frac{1}{1 + \alpha^{r}_i}.
\label{eq:re_p}
\end{equation}
Equations~\ref{eq:uni_p} and~\ref{eq:re_p} indicate that the unidirectional and reciprocated links can be considered separately.

\subsection{The maximally constrained null model}
The maximally constrained null model considers the unidirectional, reciprocated and conflicting links separately.

Randomizing signs of links at a given topology is equivalent to randomizing a subnetwork from a given network. 
For a unidirectional network, we randomize the negative subnetwork ($G_{u-}$) from the unidirectional network ($G_{u0}$) and set the rest of the links as positive. From Equations~\ref{eq:k_i_in} and~\ref{eq:k_i_out}, the parameters related to negative in- and out-degrees can be found iteratively as
\begin{equation}
    \alpha_i^{in_u-'} = \frac{1}{k_i^{in_u-}} \sum_{(i,j) \in G_{u0}} \frac{1}{1/\alpha_i^{in_u-} + \alpha_j^{out_u-} },
\end{equation}
\begin{equation}
    \alpha_i^{out_u-'} = \frac{1}{k_i^{out_u-}} \sum_{(i,j) \in G_{u0}} \frac{1}{1/\alpha_i^{out_u-} + \alpha_j^{in_u-} },
\end{equation}
where $k_i^{in_u-}$ and $k_i^{out_u-}$ are the negative in- and out-degrees in subnetwork $G_{u-}$, while the summation is over all links in the network $G_{u0}$.

The probability of a link in $G_{u0}$ to have a negative sign is given by 
\begin{equation}
    p_{ij}^{u-} = \frac{1}{\alpha_i^{in_u-} \alpha_j^{out_u-} + 1},
\end{equation}
otherwise, the link is assigned a positive sign.

For the case of mutual links, to randomize signs, we consider mutual links with the same sign and different signs separately, corresponding to the reciprocated and conflicting links.
For the reciprocated network with the same signs, it can be considered as an undirected network. We denote the undirected network and the negative subnetwork as $G_{r0}$ and $G_{r-}$, respectively.
From Equation \ref{eq:k_r}, we have
\begin{equation}
    \alpha_i^{r-'} =\frac{1}{k_i^{r-}} \sum_{(i,j) \in G_{r0}} \frac{1}{1 + \alpha_i^{r-}},
\end{equation}
where $k_i^{r-}$ is the negative reciprocated degree. Similarly, the probability for a link in $G_{r0}$ to have a negative sign is given by 
\begin{equation}
    p_{ij}^{r-} = \frac{1}{1 + \alpha_i^{r-}},
\end{equation}
and the rest of the links are set to positive signs automatically.

For conflicting links, positive out-degrees equal negative in-degrees, and positive in-degrees equal negative out-degrees. We simplify such a network by randomly keeping one link of the mutual links and randomizing them as a unidirectional network to determine the signs. Once the sign of the unidirectional link is determined, the sign of the corresponding mutual link is automatically assigned. For example, if $A\rightarrow B$ has a negative sign then the link $A\leftarrow B$ with a positive sign is added back after the randomization. 

For all parameters $\alpha$, we start with the initial value of $1$ and perform $10^4$ iterations to ensure convergence. 
After randomizing the unidirectional, reciprocated, and conflicting networks separately, they are combined to form an instance of the null model. We generate 1000 such instances for each dataset and count the triads for each instance separately.

\subsection{The signed directed null model}
The signed directed null model is a less constrained null model that focuses on considering the in- and out-degrees. Ignoring the reciprocated degree leads to 
\begin{equation}
    \alpha_i^{in-'} = \frac{1}{k_i^{in-}} \sum_{(i,j) \in G_{0}} \frac{1}{1/\alpha_i^{in-} + \alpha_j^{out-} },
\end{equation}
\begin{equation}
    \alpha_i^{out-'} = \frac{1}{k_i^{out-}} \sum_{(i,j) \in G_{0}} \frac{1}{1/\alpha_i^{out-} + \alpha_j^{in_u-} },
\end{equation}
where $k_i^{in-}$ and $k_i^{out-}$ are the negative in- and out-degrees in subnetwork $G_{-}$ with all negative links in the original network $G_0$. The probability of a link in $G_{0}$ to have a negative sign is given by 
\begin{equation}
    p_{ij}^{-} = \frac{1}{\alpha_i^{in-} \alpha_j^{out-} + 1},
\end{equation}
otherwise, the link is assigned a positive sign. We initialize $\alpha_i^{in-'}=1$ and $\alpha_i^{out-'}=1$ and perform $10^4$ iterations to ensure convergence. 1000 instances of the signed directed null model are generated for each dataset.

\FloatBarrier
\section{Results for triads with conflicting links}
\begin{figure*}[htpb]
\centering
\includegraphics[width=\linewidth]{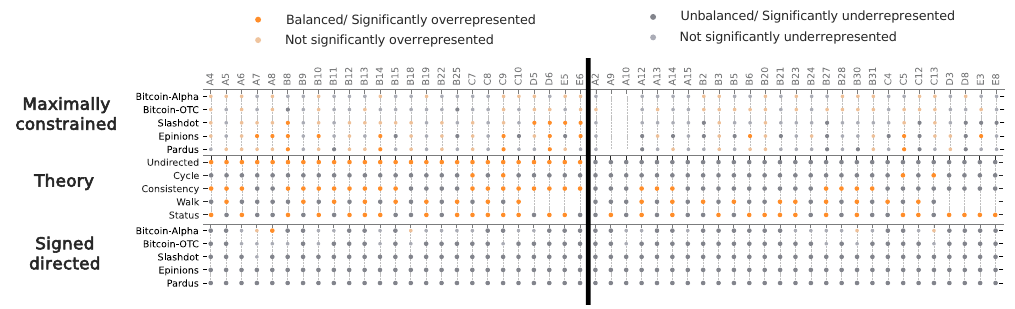}
\caption{Comparison of observed triad statistics to null models and balance theories for triads with conflicting links. 
Triads are ordered based on their balance status according to the \textit{Undirected} definition, with balanced triads presented first, followed by a black line and then unbalanced triads.
The ``Maximally constrained'' and ``Signed directed'' rows show $z$-scores quantifying the deviation of triad frequencies in empirical social networks from corresponding null model expectations. Orange (gray) dots indicate significant overrepresentation (underrepresentation) with $z>2$ ($z<-2$), while lighter colors indicate insignificant results with ($|z|<=2$). The ``Theory" rows indicate whether each triad configuration is classified as balanced (orange) or unbalanced (gray) according to a certain definition. 
The spot is left blank if the balance can not be determined for such a triad. 
}
\label{fig:results}
\end{figure*}

\FloatBarrier
\section{Robustness of results against the threshold of $z$-score.}
In our tests, 96 correlated variables are tested simultaneously, raising the issue of potential problems with multiple hypotheses testing.
Therefore, the threshold of $z$-score needs to be selected with caution. 
In the main text, we employ a threshold of $|z| = 2$ to determine whether the underrepresentation or overrepresentation is significant. To demonstrate that our main conclusions are robust against different choices of thresholds, we conduct additional analyses with varying $z$-score cutoffs.

\begin{figure}[htpb]
\centering
\includegraphics[width=\linewidth]{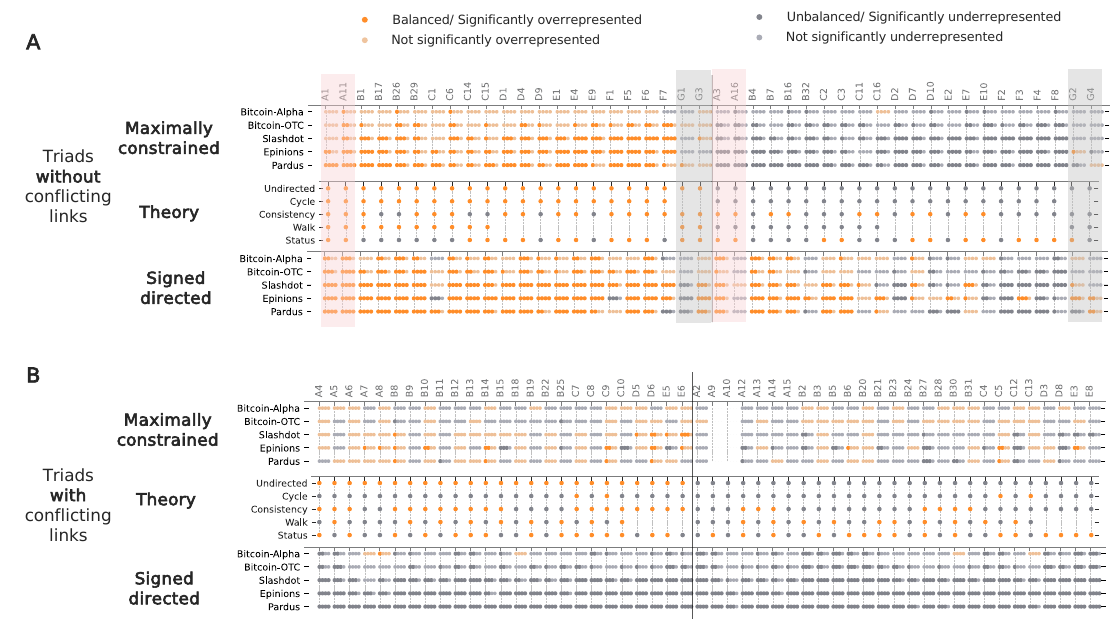}
\caption{Robustness analysis of triad representation using more stringent $z$-score thresholds. This figure illustrates the significance of over- and underrepresentation for various triad types across five networks under increasingly conservative $z$-score thresholds: $|z| = 2$ (left column, main results), $|z|=3$ (second column), $|z| = 5$ (third column), and $|z| = 10$ (last column). Panel A shows triads with non-conflicting links, while Panel B displays triads with conflicting links. Theoretical predictions of balance are included for reference.}
\label{figS:robust_z2_3_5_10}
\end{figure}

Given that multiple hypothesis testing is prone to Type I errors, which represent false positive findings, 
Increasing the $z$-score threshold for a more stringent test should reduce Type I errors, which represent false positive findings. In Figure \ref{figS:robust_z2_3_5_10}, we present the results of applying more conservative thresholds: $|z| = 2, 3, 5,$ and $10$. Notably, the overall patterns of over- and underrepresentation remain consistent across these thresholds. The persistence of key trends, even at the highly conservative threshold of $|z| = 10$, demonstrates the robustness of our primary conclusions.


\FloatBarrier
